\documentclass[aps,prab,superscriptaddress,twocolumn]{revtex4-2}

\usepackage{graphicx}
\usepackage[per-mode=symbol]{siunitx}
\usepackage{dcolumn}
\usepackage{bm}
\usepackage{booktabs}
\usepackage{xcolor}
\usepackage{multirow,ragged2e} %
\usepackage[none]{hyphenat} 
\usepackage{textcomp}
\usepackage{listings}
\usepackage{amsmath}
\usepackage{amssymb}
\usepackage{derivative}
\usepackage{mathtools} 
\usepackage[hidelinks]{hyperref}
\usepackage{subfigure}
\usepackage{tikz}
\usetikzlibrary{shapes,arrows,positioning,fit}
\usepackage[USenglish]{babel}
\usepackage{placeins}

\usepackage{nicefrac}

\newcommand{\nhdots}{\kern-0.25em\hdots\kern-0.25em}

\begin{document}

%\preprint{}

\title{Estimation of superconducting cavity bandwidth and detuning using a Luenberger observer}



\newcommand{\DESY}{Deutsches Elektronen-Synchrotron DESY, Notkestraße 85, 22607 Hamburg, Germany}
\newcommand{\TUHH}{Technische Universität Hamburg TUHH,
Am Schwarzenberg-Campus 1, 22549 Hamburg, Germany}
\newcommand{\SCKCEN}{Belgian Nuclear Research Centre SCK-CEN,
Boeretang 200, 2400 Mol, Belgium}
\newcommand{\UHH}{Universität Hamburg UHH, Mittelweg 177, 22549 Hamburg, Germany}

\author{Bozo Richter}
\email[]{bozo.richter@desy.de}
\affiliation{\DESY}
\affiliation{\TUHH}

\author{Andrea Bellandi}
\affiliation{\SCKCEN}

\author{Julien Branlard}%
\affiliation{\DESY}

\author{Leon Speidel}
\affiliation{\UHH}

\author{Annika Eichler}
\affiliation{\DESY}
\affiliation{\TUHH}


\date{\today}

\begin{abstract}
Enabled by progress in superconducting technology, several continuous wave linear accelerators are foreseen in the next decade. For these machines, it is of crucial importance to track the main cavity parameters, such as the resonator bandwidth and detuning. The bandwidth yields information on the superconducting state of the cavity. The detuning should be minimized to limit the required power to operate the cavity. The estimation of these parameters is commonly implemented in the digital electronics of the Low-Level RF control system to minimize the computation delay. In this proceeding, we present a way to compute the bandwidth and detuning using a Luenberger observer. In contrast to previous methods, a state observer yields estimations at the native control system sample rate without explicitly filtering the input signals. Additionally, the error convergence properties of the estimations can be controlled intuitively by adjusting gain parameters. Implementation considerations and test results on the derived observer are presented in the manuscript.
\end{abstract}

%\keywords{}

\maketitle

\section{Introduction}
\label{introduction}
The operation of future superconducting radio-frequency (SRF) linear particle accelerators, such as LCLS-II-HE~\cite{galayda2018lcls,raubenheimer2018lcls}, SHINE~\cite{huang2021features,liu2022progress}, and the European XFEL (EuXFEL)~\cite{sekutowicz_research_2015,vogel2019status,gjonajbeam}, relies heavily on advanced diagnostic and control methods. These methods are essential to maximize the availability and efficiency of accelerators, particularly when operating at high RF duty cycles with long pulses or in continuous wave (CW). Accurately estimating cavity detuning and bandwidth is a key requirement in achieving optimal performance for SRF particle accelerators.

Cavity detuning $\Delta\omega$ is the difference between the cavity resonance frequency $\omega_{0}$ and the driving signal frequency $\omega$, and should be minimized in order to reduce the accelerating system RF power consumption. Pressure drifts, ponderomotive instabilities and microphonic effects can increase the magnitude of detuning~\cite{bellandi_llrf_2021}.

The ratio of detuning to half bandwidth has a significant impact on power requirements. New CW facilities operate at low cavity half bandwidths in the range of tens of Hertz, i.e. comparable to magnitudes of detuning. Therefore, realtime controllers compensating the detuning are required~\cite{holzbauer_active_2018}, and they require accurate estimates of the detuning to act on.

The cavity half bandwidth $\omega_{1/2}$ is inversely proportional to the quality factor of the RF cavity resonator system, including energy dissipation in the cavity itself and in externally coupled peripherals. Sudden dissipative processes like quenches can be identified by observing the half bandwidth, increasing the operational safety and reducing potential cryogenic downtime.

%The bandwidth of the tuning disturbances for TESLA-like cavities is roughly $\SI{1}{\kilo \hertz}$ and given by the frequency of cavity mechanical resonances.
In the past there were already attempts to compute these parameters inside the field programmable gate array (FPGA) logic of low-level RF (LLRF) systems \cite{rybaniec2014real,bellandi2021online}.
The major limitation of these methods is their dependence on explicit differentiation of the cavity RF probe signal.
To limit the effect of noise on the estimation of the parameters, it is necessary to apply low-pass filters.
However, the smoothing produced by the filtering process results in signal distortions at the boundaries of pulse transitions.
Consequently, a compromise in noise reduction and pulse transition distortion has to be made.

Recent work on cavity parameter identification proposed to use the recursive least square (RLS) technique to calculate $\Delta\omega$ and $\omega_{1/2}$~\cite{PhysRevAccelBeams.26.112003}.
Using RLS might resolve the pulse boundary problem, but the algorithm needs an initial estimate of the cross-correlation matrix. 

In this paper we present an alternative technique based on the Luenberger observer~\cite{luenberger1964observing, luenberger1966observers}. This state estimation method is chosen to simplify the future FPGA implementation of the component, to maximize the sample rate, and thereby to minimize the estimation delay. Compared to the other techniques in literature, the Luenberger observer also provides filtered estimations of the cavity accelerating gradient. Finally, the error dynamics of the component can be tuned by simply defining the desired parameter estimation bandwidth through observer pole placement.

In Section~\ref{sec:cavmod}, the RF cavity model is introduced. Section~\ref{sec:derivation} is devoted to the derivation of the observer. In Section~\ref{sec:errana} the observers estimation error is analyzed formally and in simulation, and Section~\ref{sec:exp} presents results from experimental data of SRF accelerating modules at EuXFEL facilities.

\section{Cavity Model}\label{sec:cavmod}
Superconducting cavity RF dynamics without beam loading are generally described by the state space model~\cite{Schilcher:291638}
\begin{equation}\label{eqn:statespace_general}
    \odv{}{t}{\begin{bmatrix}
        v_{I}\\
        v_{Q}
    \end{bmatrix}}
    =
    \begin{bmatrix}
        -\omega_{1/2} & -\Delta\omega\\
        \Delta\omega & -\omega_{1/2}
    \end{bmatrix}\begin{bmatrix}
        v_{I}\\
        v_{Q}
    \end{bmatrix}
    +
    2\omega_{1/2}
    \begin{bmatrix}
        u_{I}\\
        u_{Q}
    \end{bmatrix} .
\end{equation}
Here, $v_I$ and $v_Q$ are the inphase and quadrature components of the RF cavity probe signal, and $u_I$ and $u_Q$ are the inphase and quadrature components of the RF cavity forward signal. In this paper, all quantities except the input coupling factor $2\omega_{1/2}$ are assumed to be time-varying.

The cavity half bandwidth reflects dissipative events in the cavity walls by its relation to the cavity unloaded quality factor $Q_0$ and is defined as
\begin{equation}\label{eqn:intro:omega12}
    \omega_{1/2} = \frac{\omega_0}{2Q_L} = \frac{\omega_0}{2Q_0} + \frac{\omega_0}{2Q_\mathrm{ext}}
\end{equation}
where $Q_\mathrm{ext}$ is the external and $Q_L$ the loaded quality factor.
During nominal operation of SRF cavities, \mbox{$Q_0 \gg Q_\mathrm{ext}$} and 
\begin{equation}\label{eqn:intro:omega12approx}
Q_L \simeq Q_\mathrm{ext} \quad \Leftrightarrow \quad \omega_{1/2} \simeq {\omega_{1/2}^\mathrm{ext} = \frac{\omega_0}{2Q_\mathrm{ext}}}~. 
\end{equation}

 When either $Q_0$ or $Q_\mathrm{ext}$ decreases, an increase of cavity bandwidth is observed, cf.~\eqref{eqn:intro:omega12}. Changes in $Q_\mathrm{ext}$ are caused by coupler movement or heating with time constants of seconds up to hours, while anomalous dissipation events lead to significant changes in $Q_0$ within fractions of a second. Hence, the effects can be distinguished clearly.

 Solving the cavity equations \eqref{eqn:statespace_general} by $\Delta\omega$ and $\omega_{1/2}$ leads to the approach of \cite{rybaniec2014real,bellandi2021online}
\begin{align}
    \begin{split}
        \omega_{1/2} &= \frac{v_I(2\omega_{1/2}^\mathrm{ext}u_I - \dot{v_I}) + v_Q(2\omega_{1/2}^\mathrm{ext}u_Q - \dot{v_Q})}{v_{\mathrm{acc}}^2}, \\
     \Delta \omega &= \frac{v_Q(2\omega_{1/2}^\mathrm{ext}u_I - \dot{v_I}) - v_I(2\omega_{1/2}^\mathrm{ext}u_Q - \dot{v_Q})}{v_{\mathrm{acc}}^2},
    \end{split} \label{eqn:intro:inverse_model}
\end{align}
where the approximation of \eqref{eqn:intro:omega12approx} is applied for the input coupling and \mbox{$v_{\mathrm{acc}} = \sqrt{v_I^2 + v_Q^2}$} is the cavity accelerating voltage.

\section{Derivation of the observer}\label{sec:derivation}
A Luenberger observer makes use of a known linear time-invariant (LTI) state space model $\mathbf{\dot{x}} = A\mathbf{x}+B\mathbf{u}$ and $\mathbf{y}=C\mathbf{x}$ to extract system state information from measured system outputs. In the state space model, $\mathbf{x}$, $\mathbf{u}$, and $\mathbf{y}$ are the vectors of system states, inputs, and outputs, and $A$, $B$, and $C$ are the system, input and output matrices. If the system representation is minimal and observable, all states can be recovered by a Luenberger observer.

Parallel to operation of the plant, the model is simulated using the same inputs as applied to the plant. The difference between measurements of plant outputs and simulated outputs is multiplied by an observer gain matrix $L$ to correct the simulation models internal states~$\mathbf{\hat{x}}$. This completes the structure of a Luenberger observer, as shown in Fig.~\ref{fig:block_diagram}.

If the model describes the plant dynamics perfectly, the difference between the observers state dynamics 
\begin{equation}\label{eqn:observer_state_dynamics}
    \mathbf{\dot{\hat{x}}}=A\mathbf{\hat{x}}+B\mathbf{u} + L(\mathbf{C\hat{x}} - \mathbf{y})
\end{equation}
and the model dynamics yields the estimation error dynamics
\begin{equation}\label{eqn:esterrdyn}
    \mathbf{\dot{x}}-\mathbf{\dot{\hat{x}}} = \mathbf{\dot{e}} = (A+LC)\mathbf{e}~.
\end{equation}
These dynamics represent an autonomous system, as its states are independent of any input. By careful selection of the entries of $L$, the eigenvalues of $A+LC$ can be placed at values leading to desired error dynamics.

However, in the general cavity state space form~\eqref{eqn:statespace_general}, the state vector contains neither $\Delta\omega$ nor $\omega_{1/2}$. Thus, the system representation is transformed assuming that significant variation in cavity half bandwidth occurs only by (partial) loss of superconductivity, i.e. greatly reduced $Q_0$. Hence, \eqref{eqn:intro:omega12} is expressed as
\begin{equation}\label{eqn:derivation:omega12}
    \omega_{1/2}(t) = \omega_{1/2}^{\mathrm{ext}} + \Delta\omega_{1/2}(t),
\end{equation}
collecting dynamic deviations of half bandwidth in the excess half bandwidth $\Delta\omega_{1/2}$. As in~\eqref{eqn:intro:inverse_model}, the input coupling factor is assumed to be constant. Then, by definition of 
\begin{equation}
    V(\mathbf{x}) = \begin{bmatrix}
        -v_I & -v_Q \\
        -v_Q & v_I
    \end{bmatrix}\quad 
    \mathbf{I} = \begin{bmatrix}
        1 & 0 \\ 0 & 1
    \end{bmatrix}\quad
    \mathbf{0} = \begin{bmatrix}
        0 & 0 \\ 0 & 0
    \end{bmatrix}
\end{equation}
and with \eqref{eqn:derivation:omega12}, one can extend the state vector of \eqref{eqn:statespace_general} and reformulate the state space system to
\begin{equation}\label{eqn:statespace_augmented}
    \begin{alignedat}{2}
       \mathbf{\dot{x}} &= A(\mathbf{x})\mathbf{x} + B\mathbf{u} &\qquad
       \mathbf{y} &= C \mathbf{x}\\
       \mathbf{x}^{\top} &= \begin{bmatrix}
            v_{I}&
            v_{Q}&
            \Delta\omega_{1/2}&
            \Delta\omega
       \end{bmatrix}
       &\quad
       \mathbf{u}^{\top} &= \begin{bmatrix}
           u_{I}&
           u_{Q}
       \end{bmatrix}
       \\
        A(\mathbf{x})&=\begin{bmatrix}
            - \omega_{1/2}^{\mathrm{ext}}\mathbf{I} & V(\mathbf{x})\\
            \mathbf{0} & \mathbf{0}
        \end{bmatrix}&
        B&=2\omega_{1/2}^{\mathrm{ext}} \begin{bmatrix}
            \mathbf{I} \\
            \mathbf{0}
        \end{bmatrix}\\
        C &= \begin{bmatrix}
            \mathbf{I} & \mathbf{0}
        \end{bmatrix}. && 
    \end{alignedat}
\end{equation}
As the system matrix $A$ is now dependent of elements of the state vector $\mathbf{x}$, \eqref{eqn:statespace_augmented} is not an LTI, but a quasi-linear parameter varying (qLPV) state space model~\cite{balas_linear_2002}. Variation of the external half bandwidth due to drifts in RF transmission line or coupler parameters is neglected throughout this work. Thus, the input and output matrices $B$ and $C$ are assumed to be constant.

\subsection*{Linear time-varying Luenberger Observer}

Using pole placement, a Luenberger observer (LO) with time-varying gain matrix $L$ is designed for the qLPV model~\eqref{eqn:statespace_augmented}. To simplify the derivation, derivatives of the time-varying parameters $v_I$ and $v_Q$ are neglected. Hence, \eqref{eqn:statespace_augmented} is interpreted as an LTI model with slow drifting parameters. The observer gain $L$ is adjusted to compensate the parameter dependence of $A$, resulting in time invariant error dynamics. A block diagram of the LO is shown in Fig.~\ref{fig:block_diagram}, the algebraic derivation of $L$ is displayed in Appendix~\ref{app:LTV_observer}.

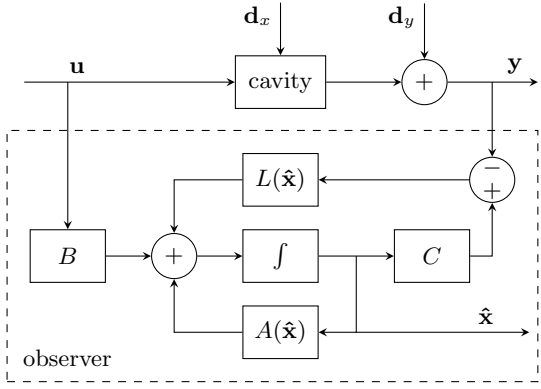
\begin{figure}
    \begin{tikzpicture}
    \def\boxheight{0.7cm}
    \def\circdia{0.6cm}
    % cavity
    \node [draw,
    	minimum width=1.2cm, 
    	minimum height=\boxheight,
    ] (cavity) at (0, 0) {cavity};
    % integrator
    \node [draw,
    	minimum width=1cm, 
    	minimum height=\boxheight, 
    	below = 1.6cm of cavity
    ]  (integrator) {$\int$};
    % Sum shape BLA
    \node[draw,
    	circle,
    	minimum size=\circdia,
    	left = 0.6cm of integrator
    ] (dynsum){};
    \node at (dynsum.center){$+$};
    % Sum shape dy
    \node[draw,
    	circle,
    	minimum size=\circdia,
    	right = 1cm of cavity
    ] (meassum){};
    \node at (meassum.center){$+$};
    % B
    \node [draw,
    	minimum width=1cm,
    	minimum height=\boxheight,
    	left=0.6cm of dynsum
    ]  (B) {$B$};
    % C
    \node [draw,
    	minimum width=1cm, 
    	minimum height=\boxheight, 
    	right= 1cm of integrator
    ]  (C) {$C$};
    % A
    \node [draw,
    	minimum width=1cm, 
    	minimum height=\boxheight, 
    	below= 0.3cm of integrator
    ]  (A) {$A(\mathbf{\hat{x}})$};
    % L
    \node [draw,
    	minimum width=1cm, 
    	minimum height=\boxheight, 
    	above= 0.3cm of integrator
    ]  (L) {$L(\mathbf{\hat{x}})$};
    % Sum shape 2
    \node[draw,
    	circle,
    	minimum size=\circdia,
    	right=2cm of L
    ] (errsum){};
    \node[above=-2pt] at (errsum.center){\small $-$};
    \node[below=-2pt] at (errsum.center){\small $+$};
    %
    % Arrows with text label
    \draw[stealth-] (cavity.west) -- ++(-2.8,0) 
    	node[near end, above]{$\mathbf{u}$};
    \draw[stealth-] (cavity.north) -- ++(0,0.7) 
    	node[near end, left]{$\mathbf{d}_x$};
    \draw[stealth-] (meassum.north) -- ++(0,0.74) 
    	node[near end, left]{$\mathbf{d}_y$};
    \draw[-stealth] (meassum.east) -- ++(1.2, 0)
    	node[near end,above]{$\mathbf{y}$};
    \draw[-stealth] (A.east)+(0.5, 0) -- ++(2.8, 0)
        node[near end, above] {$\mathbf{\hat{x}}$};
    % arrows
    \draw[-stealth] (B |- cavity) -- (B.north);
    \draw[-stealth] (cavity.east) -- (meassum.west);
    % vert/hor intersection of B and cavity
    \draw[-stealth] (B.east) -- (dynsum.west);
    \draw[-stealth] (dynsum.east) -- (integrator.west);
    \draw[-stealth] (integrator.east) -- (C.west); 
    \draw[-stealth] (C.east) -| (errsum.south);
    \draw[-stealth] (errsum.west) -- (L.east);
    \draw[-stealth] (L.west) -| (dynsum.north);
    \draw[-stealth] (A.west) -| (dynsum.south);
    \draw[-stealth] (integrator.east)+(0.5, 0) |- (A.east);
    \draw[-stealth] (errsum |- cavity) -- (errsum.north);
    % box
    \node[draw, thin, dashed, fit=(integrator) (dynsum) (B) (C) (A) (L) (errsum), 
      inner sep=0.3cm, 
      %label=below left:{observer}
      ] (observerbox) {};
    \node[anchor=south west, font=\small] at ([xshift=3pt, yshift=3pt]observerbox.south west) {observer};
    \end{tikzpicture}
    
    \caption{Block diagram of the Luenberger observer.}
    \label{fig:block_diagram}
\end{figure}

All four eigenvalues of the estimation error dynamics~\eqref{eqn:esterrdyn} are time invariant and equal to $p\in\mathbb{R}$ when applying the gain matrix
\begin{equation}\label{eqn:observer_gain_cont}
    L(\mathbf{x}) =
    \begin{bmatrix}
        (2p + \omega_{1/2}^\mathrm{ext})\mathbf{I}\\
        \dfrac{-p^2}{v_\mathrm{acc}^{2}}V(\mathbf{x})
    \end{bmatrix}.
\end{equation}
The magnitude of the time-varying entries is bounded by $p^2/v_\mathrm{acc}$, since with $i \in \{1,2\}$
\begin{equation}
    \left\vert\dfrac{x_i}{v_\mathrm{acc}^{2}}\right\vert \le \dfrac{1}{v_\mathrm{acc}}~.
\end{equation}
Hence, the time-varying adaption gains of $\Delta\omega_{1/2}$ and $\Delta\omega$ scale inversely proportional up to $v_\mathrm{acc}$, and a necessary condition for application of the LO is $v_\mathrm{acc} > 0$.

\subsection*{Discretization}
For application in sampled data systems, exact discretization is used to obtain a discrete time representation of the cavity model \eqref{eqn:statespace_augmented}. Constant inputs are assumed between sampling instances $k$. A sampling instance $k$ is related by the sampling time $T$ to continuous time $t=kT$. The discretized system and input matrices $\Phi$ and $\Gamma$ are given in block form by
\begin{equation}
    \begin{alignedat}{2}
        \Phi_k(T) &= 
        \begin{bmatrix}
            (1-\alpha)\mathbf{I} & \frac{\alpha}{w} V_k\\
            \mathbf{0}&\mathbf{I}
        \end{bmatrix}
        &\quad 
        \Gamma(T) &=2\alpha\begin{bmatrix}
            \mathbf{I} \\ \mathbf{0}
        \end{bmatrix}
        \\[1em]
        V_k&= V(\mathbf{x}_k)
        &
        \alpha = 1-&e^{-\omega_{1/2}^{\mathrm{ext}}T},
    \end{alignedat}
\end{equation}
as shown in Appendix~\ref{app:discretization}.

For the discretized system, a state dependent LO gain matrix $\Lambda_k$ is designed, as in the continuous case. All four time invariant eigenvalues of the estimation error dynamics are placed in $\rho\in\mathbb{R}$ by application of 
\begin{equation}\label{eqn:observer_gain_disc}
    \Lambda_k = \begin{bmatrix}
        (\alpha-1+2\rho-1)\mathbf{I}\\
        -\frac{\mu}{v_\mathrm{acc}^2} V_k
    \end{bmatrix}
\end{equation}
where
\begin{equation}
    \mu = -\dfrac{\omega_{1/2}^{\mathrm{ext}}}{\alpha}\left(\rho-1\right)^2.
\end{equation}
To obtain discrete time dynamics equivalent to continuous time poles in $p$, the eigenvalue mapping \mbox{$\rho = e^{pT}$} can be applied. Restrictions in the choice of $p$ and the derivation of $\Lambda_k$ are given in Appendix~\ref{app:discretePoles}.

In the presented LO implementation, the estimation error decay bandwidth defined by the pole $p$ remains as a free design parameter.
%bandwidth gain factor $f_1$, and the detuning gain factor $f_2$.
Nevertheless, the external half bandwidth $\omega_{1/2}^{\mathrm{ext}}$ and an initial observer state vector $\mathbf{x}_0$ must be provided for application.

\section{Error analysis}\label{sec:errana}
To study the effects of process and output disturbances on the estimation error, two inputs $\mathbf{d}_x$ and $\mathbf{d}_y$ (see Fig.~\ref{fig:block_diagram}) are added to the qLPV cavity model \eqref{eqn:statespace_augmented}. A filter block \begin{equation}
    H(s) = \begin{bmatrix}
        \mathbf{I} & \mathbf{0} \\
        \mathbf{0} & s\mathbf{I}
    \end{bmatrix}
\end{equation} is applied to $\mathbf{d}_x$ to make its bottom entries correspond to $\Delta\omega_{1/2}$ and $\Delta\omega$ of the true cavity system.

Given exact knowledge of the external half bandwidth and RF states to formulate the LO matrices, the continuous time estimation dynamics are 
\begin{equation}\label{eqn:error_dynamics}
    \mathbf{\dot{\hat{x}}} =A(\mathbf{x})\mathbf{\hat x} + B\mathbf{u} - L(\mathbf{x})\big(C\mathbf{e} + d_y\big)
\end{equation}
leading to the error dynamics
\begin{align}\label{eqn:derivation:errordynamics}
\begin{split}
    \mathbf{\dot{e}} = \big(A&(\mathbf{x}) + L(\mathbf{x})C\big)\mathbf{e}
    + H\mathbf{d}_x + L\mathbf{d}_y.
\end{split}
\end{align}
Application of the Laplace transformation and rearraning the equation yields the estimation error transfer matrices from the error driving terms $\mathbf{d}_x$ and $\mathbf{d}_y$
\begin{align}\label{eqn:tfmatrices}
\begin{split}
    G_{ed_x} &= \left(s\mathbf{I}_4-A(\mathbf{x})-L(\mathbf{x})C\right)^{-1}H\\
    G_{ed_y} &= \left(s\mathbf{I}_4-A(\mathbf{x})-L(\mathbf{x})C\right)^{-1}L
    \end{split}
\end{align}
where $\mathbf{I}_4$ is the $4\times 4$ identity matrix.
The tracking dynamics for half bandwidth and detuning can be characterized by the right block of the transfer matrix
\begin{equation}\label{eqn:analysis:tracking}
    \begin{bmatrix}
        % \mathbf{I} & \mathbf{0} \\
        \mathbf{0} & \mathbf{I}
    \end{bmatrix}(\mathbf{I}_4- G_{ed_x})
    = 
    \dfrac{p^2}{(s-p)^2}\begin{bmatrix}
        % \left((s-p)^2+s)\right)\mathbf{I} & (s-p)^2 - \mathbf{V}(\mathbf{x})\\
        \dfrac{1}{v_\mathrm{acc}^2}V(\mathbf{x}) & \mathbf{I}
        % \dfrac{1}{v_\mathrm{acc}}\mathbf{[1]} & \mathbf{I}
    \end{bmatrix}.
\end{equation}
Consequently, the LO incorporates a decoupled pair of second order low pass filters acting on the true $\Delta\omega_{1/2}^\mathrm{ext}$ and $\Delta\omega$ of the cavity system, given an error free setup of LO matrices.

The left block of~\eqref{eqn:analysis:tracking} reveals that the parameter estimates are perturbed by process disturbances acting on $v_I$ and $v_Q$, scaled by up to $\vert 1/v_{\mathrm{acc}}\vert$. These RF process disturbances are also second order low-pass filtered with bandwidth $\vert p\vert$ and have a nonzero static gain, thus they propagate to the estimates of $\Delta\omega_{1/2}$ and $\Delta\omega$ in steady state. They represent any input discrepancies between cavity and LO, e.g. input measurement errors, beam effects, or incorrectly chosen external half bandwidth.

The transfer characteristics of output measurement disturbances $\mathbf{d}_y$ to estimation errors of half bandwidth and detuning are given by
\begin{equation}
    \begin{bmatrix}
        \mathbf{0} & \mathbf{I} 
    \end{bmatrix} G_{ed_y}
    =\dfrac{p^2(s+\omega_{1/2}^{\mathrm{ext}})}{(s-p)^2}\cdot\dfrac{1}{v_\mathrm{acc}^2}V(\mathbf{x}).
\end{equation}
As the previously discussed RF disturbances in $\mathbf{d}_x$, these transfer functions have nonzero static gains, so static components in $\mathbf{d}_y$ lead to constant estimation errors in half bandwidth and detuning.

In application, the external half bandwidth is not known exactly, but an approximate value with an uncertainty factor $\kappa\approx 1$ in
\begin{equation}
    \hat\omega_{1/2}^\mathrm{ext}=\kappa\omega_{1/2}^\mathrm{ext}
\end{equation}
is used to setup the LO matrices $\hat A$, $\hat B$ and $\hat L$, applied in~\eqref{eqn:error_dynamics}. To describe the effect on the estimates $\Delta\hat\omega_{1/2}$ and $\Delta\hat\omega$, the steady state solution of~\eqref{eqn:error_dynamics} using $\hat A$, $\hat B$, and $\hat L$ is compared to the steady state solution of~\eqref{eqn:statespace_augmented} given the same input vector $\mathbf{u}$. For $\mathbf{d}_x=0$ and $\mathbf{d}_y=0$ one obtains the steady state estimates
\begin{align}
    \Delta\hat\omega_{1/2ss} &= \kappa\Delta\omega_{1/2ss}\\\Delta\hat\omega_{ss} &= \kappa\Delta\omega_{ss}.
\end{align}
Evaluating~\eqref{eqn:derivation:omega12} to obtain the estimated cavity half bandwidth
\begin{equation}
    \hat\omega_{1/2ss} = \kappa(\omega_{1/2ss} + \Delta\omega_{1/2ss})
\end{equation}
reveals that both parameter estimates are scaled by~$\kappa$. 

Acceleration of a charged particle beam transfers energy from the cavity to the beam. This energy loss can be compensated by a counteracting increase in forward power to maintain the cavity in steady state. However, these beam effects are not included in the observer design, hence acting as a disturbance on the estimation. The corresponding apparent change in detuning and half bandwidth due to beam effects can be calculated as shown in Appendix~\ref{sec:appendix:beam}.

\subsection*{Nonlinear Simulation}
To validate the results of the simplified linear analysis, the discrete time LO is applied in an open-loop cavity simulation. The simulation is based on the continuous time nonlinear cavity dynamics \eqref{eqn:statespace_general} evaluated by a fourth-order Runge-Kutta solver at \SI{9}{\mega\hertz} sampling rate. The detuning dynamics are simulated by a fifth-order filter excited by $v_\mathrm{acc}^2$ and an additional free input representing a mechanical actuator.

The cavity parameters are based on EuXFEL specifications, however the RF setpoint timing and scaling are adjusted to allow for clearer visualization. The external half bandwidth is set to \SI{141}{\hertz}, corresponding to $Q_L=4.6\times10^6$ at a generator frequency of \SI{1.3}{\giga\hertz}. The electro-mechanical coupling is \SI{-1}{\hertz\per(\mega\volt)^2}, and the simulated maximum $v_\mathrm{acc}$ is \SI{8}{\mega\volt}.

The mechanical actuator excites the detuning dynamics with a sinusoidal stimulus about \SI{6}{\milli\second} prior to the RF drive pulse. The RF drive pulse is composed of two sections of constant drive power: First, high power is provided in a filling phase of \SI{750}{\micro\second} where the cavity field rises to its target value for acceleration. Then follows a \SI{850}{\micro\second} flattop, where the drive power is set to maintain the cavity field. Finally, the drive is disabled, resulting in field decay. 

An artificial beam compensation term is superimposed on the simulated RF drive signal starting at \SI{1000}{\micro\second} and ending at \SI{1250}{\micro\second}.
The apparent changes in half bandwidth and detuning are \SI{14.1}{\hertz} and \SI{20}{deg}. The simulated cavity dynamics are not altered, since the additional drive term is assumed to perfectly compensate beam loading effects.

\begin{figure}[tb]
    \centering
    \includegraphics[width=8.5cm]{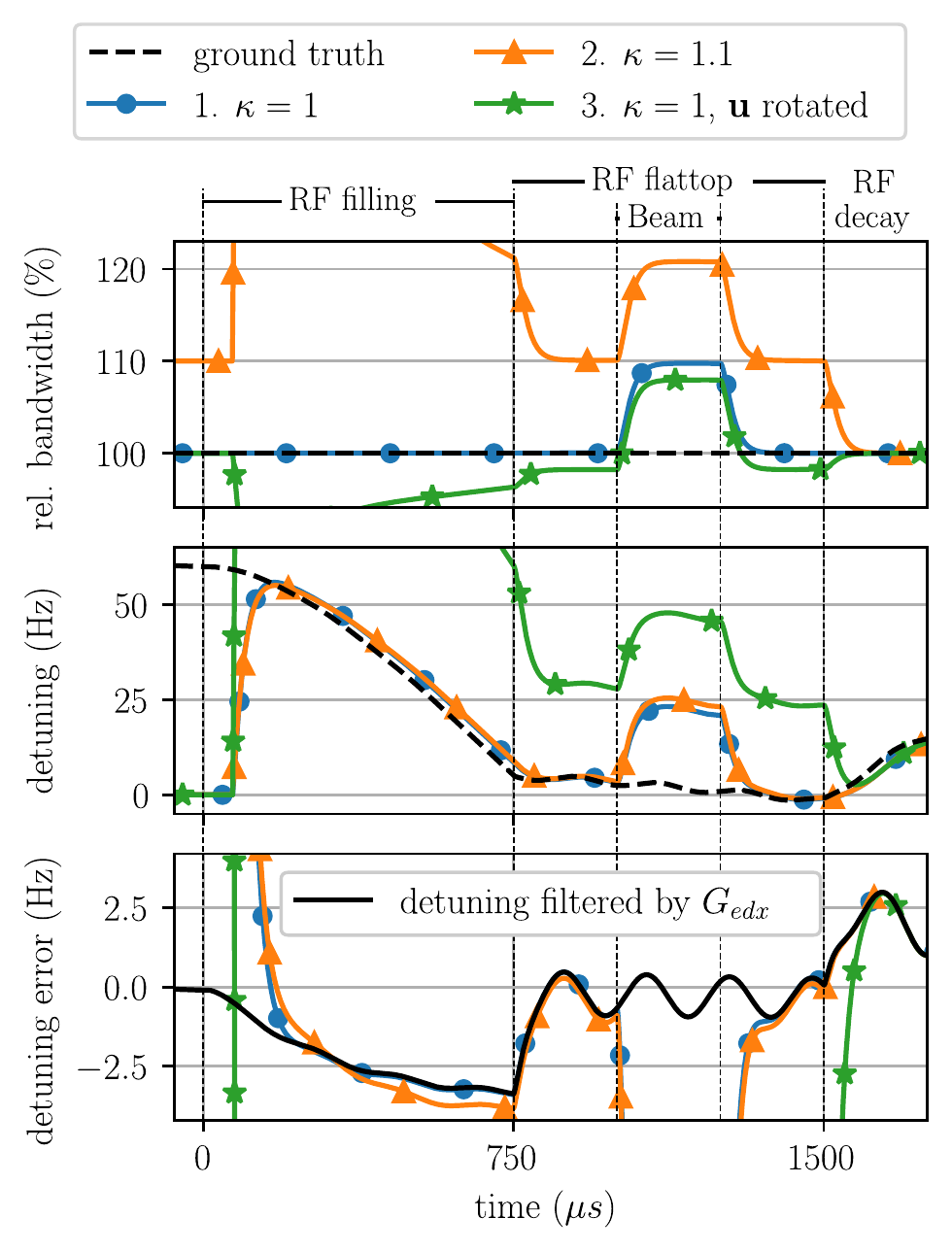}
    \caption{Estimated half bandwidth (top), detuning (center), and detuning estimation error (bottom) of a simulated RF pulse with typical XFEL cavity parameters. Simulation state trajectories (dashed) are compared to LO estimations for ideal (filled circles) and mismatched (triangles) initial conditions, and a wrongly calibrated RF drive signal (stars).}
    \label{fig:sim}
\end{figure}

After simulation and drive signal modification, the RF signals are provided to the discrete time LO, which is set up with a pole frequency of \SI{10}{\kilo\hertz}. To avoid large gains during the filling phase in the bottom rows of $\Lambda_k$, $g_{2,k}$ is set to zero until $v_\mathrm{acc}>\SI{1}{\mega\volt}$. The effects of initialization and calibration errors are investigated by setting the initial LO states to $\hat{\mathbf{x}}_0 = \mathbf{0}$ and furthermore
\begin{enumerate}
    \item $\kappa=1$
    \item $\kappa=1.1$
    \item $\kappa=1$ and $\mathbf{u}$ rotated by \SI{10}{deg}.
\end{enumerate}
The LO estimations resulting from the given scenarios are shown in Fig.~\ref{fig:sim}.

% bandwidth estimation
Scenario 1 shows a flat bandwidth estimation and tracks the detuning as expected, with constant offsets in the beam region. There, the expected apparent changes in estimated quantities is observed, with critical damping in the transitions. The detuning estimation error matches the response of $G_{dx}$ from~\eqref{eqn:tfmatrices}.

Scenario 2 and 3 both show steady state bandwidth estimation erros with strongly damped transients at drive power changes. During RF filling, the half bandwidth estimation error decay is significantly slower than during the flattop, possibly due to neglecting derivatives of RF states during LO design.

The initial estimation error of the detuning decays equally fast in Scenario 1 and 2. In Scenario 3, both bandwidth and detuning are subject to steady state estimation errors as long as drive power is supplied to the cavity. In all Scenarios, both bandwidth and detuning estimation converge correctly during RF decay, given no static offsets in RF drive and cavity measurements.

\FloatBarrier

% implications for CW
Consequently, drive power calibration error and wrong external half bandwidth initialization are clearly recognizable in pulsed operation, where the forward power varies significantly and regularly. 
During CW operation, RF drive is supplied to the cavity persistently, and a free cavity response is not available. However, if the RF drive calibration is trusted, $\omega_{1/2}^{\mathrm{ext}}$ may be corrected by introducing RF drive variations and minimizing the resulting estimated half bandwidth deviation. Otherwise, techniques like \cite{CWQL} may be required to verify the external half bandwidth and correct signal calibration, or CW operation is interrupted for calibration and computation of $\hat\omega_{1/2}^\mathrm{ext}$. 

\section{Experimental results}\label{sec:exp}
The presented discrete time LO is applied to nominal and anomalous experimental data obtained from EuXFEL and the cryo module test bench (CMTB). The data used is chosen to showcase the LO's ability to detect quenches and distinguish them from nominal pulses. At both facilities, the same type of cryomodule with SRF TESLA cavities is installed, but driven in different operating modes.

\subsection{Short Pulse}
The EuXFEL is a user facility driven by a linear accelerator, yielding an electron beam with energies up to \SI{17.5}{\giga e\volt}. The high energy electron beam is used to generate ultra short X-ray flashes with photon energies up to \SI{25}{\kilo e\volt} to enable photon science experiments. The accelerator is operated in pulsed mode at \SI{10}{\hertz} repetition rate with \SI{1.4}{\percent} RF duty factor, called short pulse (SP) mode. Each RF station consists of 4 cryomodules and a total of 32 cavities. All cavities are driven in parallel by a single klystron under vector-sum control. 

Front end CPUs have access to RF data traces with a length of $2^{14}$ samples obtained at a rate of \SI{9}{\mega\hertz}, yielding a covered time span of \SI{1.8}{\milli\second} per pulse. On occurrence of a quench, an archiving system stores decimated data at \SI{1}{\mega\hertz} for post-mortem analysis, including 200 pulses prior to and 50 more pulses beginning with the anomaly. One of these datasets is used for application of the LO. The recorded data is preprocessed by applying the energy based crosstalk removal and signal calibration method to the first 100 pulses as presented in~\cite{bellandi_calibration_2024, brandt_development_2007}.

\begin{figure}[t!]
    \centering
    \includegraphics[width=8.5cm]{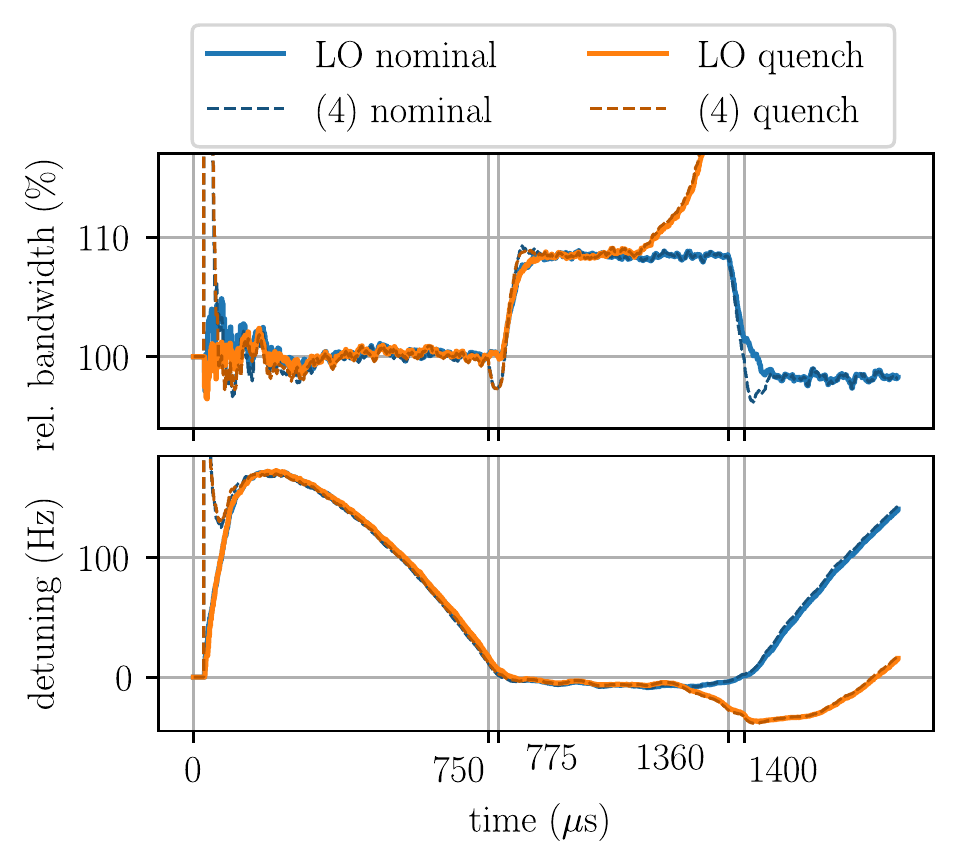}
    \caption{Bandwidth (top) and detuning (bottom) estimations of a nominal EuXFEL pulse with on-crest beam, and during a quench. For comparison, results using \eqref{eqn:intro:inverse_model} are shown.}
    \label{fig:XFEL}
\end{figure}

In Fig.~\ref{fig:XFEL}, LO estimations obtained from the last nominal and the subsequent quenching pulse are shown. The observer poles are placed at \SI{10}{\kilo\hertz}. Half bandwidth and detuning estimation are inhibited at acceleration voltages below \SI{1}{\mega\volt}. The flattop acceleration voltage is \SI{21}{\mega\volt}. During filling, the decreasing RF measurement signal to noise ratio (SNR) leads to reducing estimation noise in half bandwidth estimation. The critically damped low-pass characteristic of the estimation is observable from the beam induced apparent bandwidth change. Within the beam region, the quench is clearly visible by the sudden increase in excess bandwidth, allowing to react on an anomaly within the current pulse.

Figure~\ref{fig:XFEL} also shows estimations using the inverse model approach \eqref{eqn:intro:inverse_model}, where derivatives are obtained through second order Butterworth filtering at \SI{10}{\kilo\hertz} and subsequent first order differences calculation. Different from~\cite{bellandi2021online}, the available data is not decimated to \SI{100}{\kilo\hertz}. Notable deviations from LO results appear in the beginning of the filling phase, where the inverse approach leads to large estimation errors, and in pulse transitions, where under- and overshoot in the bandwidth estimation are observed due to distortions caused by filtering.

\FloatBarrier

\subsection{Continuous Wave}
CMTB is a test facility that can host one EuXFEL-type cryomodule with eight cavities. It is equipped with distinct LLRF systems for generator driven vector-sum and single cavity control, and self excited loop (SEL) operation. An inductive output tube provides RF drive power of up to \SI{40}{\kilo\watt} to a waveguide distribution system connecting all eight cavities. This amplifier can be operated at any duty factor, including continuous wave. At CMTB, no electron beam is available.

The front end CPU allows configurable data acquisition sample rates by decimating the available \SI{9}{\mega\hertz} sample stream of the digital signal processor (DSP). In CW operation, data is acquired at \SI{16.4}{\kilo\hertz}. An experiment is conducted using a single cavity digital SEL controller with an output amplitude limiter. The amplitude limit is increased successively, until a quench occurs at a cavity probe voltage of \SI{18}{\mega\volt}.

\begin{figure}[tb!]
    \centering
    \includegraphics[width=8.5cm]{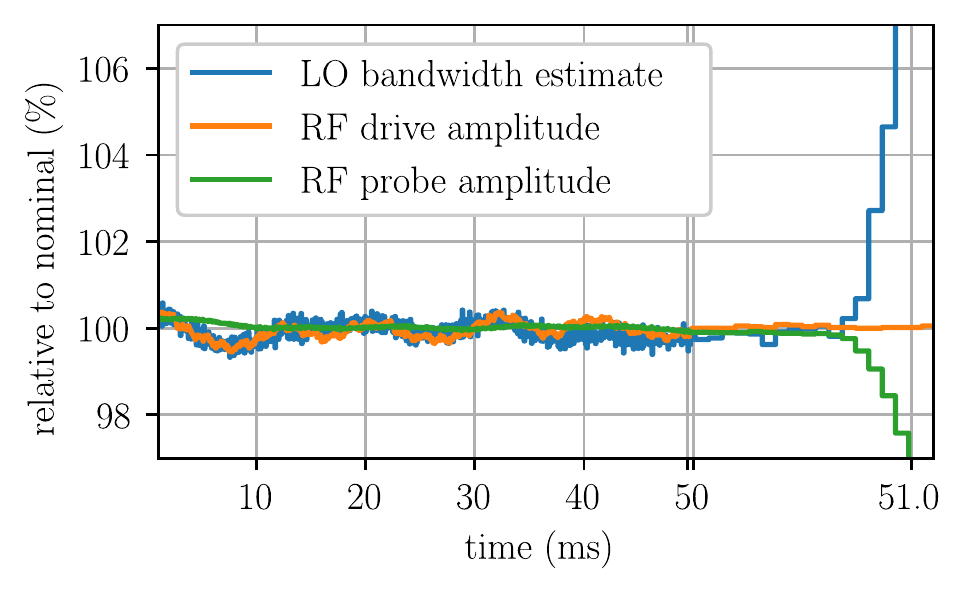}
    \caption{Normalized bandwidth estimation and cavity voltage measurements of CW quench test in SEL at CMTB.}
    \label{fig:CMTB}
\end{figure}

In Fig.~\ref{fig:CMTB}, the relative changes of $v_\mathrm{acc}$, $\vert \mathbf{u} \vert$ and the estimated half bandwidth of an LO with poles at \SI{3}{\kilo\hertz} are displayed. Starting at \SI{50}{\milli\second}, the time scale is stretched by a factor of 20. The drive power remains close to an arbitrary nominal value at all times. The cavity probe voltage decays starting at \SI{50.7}{\milli\second}, indicating a quench. The relative increase in estimated half bandwidth is notably steeper than the decrease in cavity voltage, enabling a faster reaction.

In the presented case, the quench is clearly visible from the cavity voltage alone. However, setpoint tracking controllers, detuning and beam loading pose additional challenges to implement a reliable quench detector based on direct interpretation of RF trajectories. In contrast, the LO yields an explicit estimate of the external half bandwidth in any case.

\section*{Conclusion}
An observer has been developed incorporating the nonlinear coupling of detuning and RF cavity states, producing second-order filtered time series of the cavity system detuning and half bandwidth. The method relies on accurately calibrated, offset-free RF signals and precise estimates of the cavity half bandwidth.

Compared to existing approaches, the presented observer design respects the cavity system inherent nonlinearity and offers a straightforward setup, requiring intuitive tuning and initial parameters. Application on recorded experimental data is promising. A real-time implementation operating at the native DSP sample rate remains to be tested.

% \subsection*{Future Work}
Future efforts will include quench detection based on LO estimates, and the low-delay, native-speed detuning estimation enables the implementation of a detuning-aware feedback and feedforward RF controller. Resonance control using downsampled detuning data is also anticipated.
Further exploration may include SNR evaluation at $\omega_{1/2}^{\mathrm{ext}} < \SI{65}{\hertz}$, beam parameter estimation, CW calibration, and online identification of the external half bandwidth.

\begin{acknowledgments}
This work was partially funded in the context of the R\&D program of the European XFEL. This project has received funding from the European Unions Horizon Europe research and innovation program under grant agreement N.~101131435.
\end{acknowledgments}

%\section*{Author Contributions}
%“A.B. and C.D. contributed equally to this work. A.Z. and B.Y. conceptualized the work; C.X. conducted the experiments.”

%For guidance on contributor roles, see the CRediT taxonomy.

%(As an option, the statement “These authors contributed equally to this work” may be set as a byline footnote to the relevant authors.)

\appendix
\section{Observer Pole Placement} \label{app:LTV_observer}
Derivatives of the RF states are neglected to allow for a straight forward observer design.
In the following, an observer gain
\begin{equation}
    L(x) = \begin{bmatrix}
        (\omega_{1/2}^{\mathrm{ext}} + \gamma) \mathbf{I} \\
        W
    \end{bmatrix}
\end{equation}
will be designed, leading to the estimation error dynamic matrix
\begin{equation}
    A_{\mathrm{cl}}(x) = A(x) + L(x) C = \begin{bmatrix}
        \gamma \mathbf{I} & V \\
        W & 0
    \end{bmatrix}.
\end{equation}

The chosen structure of $L$ reduces the amount of free design parameters to three and allows straight forward pole placement.
Therefore, the eigenvalues of $A_{\mathrm{cl}}$ have to be calculated. The characteristic polynomial
\begin{equation}
    \det(\lambda\mathbf{I}_4 - A_{\mathrm{cl}}) = \det\left(
    \begin{bmatrix}
        (\lambda-\gamma)\mathbf{I} & -V \\
        -W & \lambda \mathbf{I}
    \end{bmatrix}\right)
\end{equation}
is simplified using the Schur complement
\begin{equation}
    \begin{bmatrix}
        A&B\\C&D
    \end{bmatrix} = \begin{bmatrix}
        A-BD^{-1}C & B\\0&D
    \end{bmatrix}
\end{equation}
revealing that
\begin{align}\label{eqn:charpol}
    \begin{split}
        \det(\lambda\mathbf{I}_4 - A_{\mathrm{cl}}) &= \det((\lambda-\gamma)\mathbf{I}-\dfrac{1}{\lambda}VW)\det(\lambda\mathbf{I})\\
    &= \det((\lambda^2-\lambda\gamma)\mathbf{I}-VW)\\
    &= \det(\mu\mathbf{I}-VW)
    \end{split}
\end{align}
is the characteristic polynomial of $VW$ in $\mu$, requiring $\lambda \neq 0$. Assuming that \mbox{$v_\mathrm{acc}^{2} = v_I^{2}+v_Q^{2} > 0$}, static eigenvalues $\mu_i$ can be assigned by setting
\begin{align}\label{eqn:W_mu}
    W &= MV^{-1} = \begin{bmatrix}
        \mu_1 & 0\\
        0 & \mu_2
    \end{bmatrix} V^{-1} = \dfrac{1}{v_{\mathrm{acc}}^2}\begin{bmatrix}
        \mu_1 & 0\\
        0 & \mu_2
    \end{bmatrix}V
\end{align}
as $V$ is a scaled involution. Hence, $VW = VMV^{-1}$ is a similarity transformation of $M$ and thus maintains the eigenvalues $\mu_i$. The roots of \eqref{eqn:charpol} are found by solving
\begin{align}
    0 &= \lambda^{2} - \lambda\gamma - \mu_i\\
    \Rightarrow \quad \lambda_j &= \dfrac{\gamma}{2} \pm \sqrt{\dfrac{\gamma^{2}}{4} + \mu_i}
\end{align}
with $i \in \{1, 2\}$ and $j \in \{1,2,3,4\}$.

For a more intuitive observer tuning, one can introduce the pole location $p$ and additional tuning parameters $f_i\in \mathbb{R}$, and set
\begin{align}\label{eqn:conttimeeigs}
    \begin{split}
        \gamma &= 2p\\
    \mu_i &= -f_i p^2\\
    \Rightarrow\quad\lambda_j&=p \left(1 \pm \sqrt{1-f_j}\right).
    \end{split}
\end{align}
With $f_i = 1$ all $\lambda_j=p$, so the observers estimation error dynamics are critically damped. The choice of $W$ allows association of $f_1$ and $f_2$ to the speed of estimation of $\Delta\omega_{1/2}$ and $\Delta\omega$, respectively. For stable observer dynamics, one has to select $p < 0$ and $f_i > 0$ and ensure that $v_\mathrm{acc}^{2} > 0$.

Then, the parameter varying observer gain providing time invariant observer dynamics while neglecting parameter derivatives is
\begin{equation}
    L(x) = \begin{bmatrix}
        \omega_{1/2}^{\mathrm{ext}} + 2p & 0\\
        0 & \omega_{1/2}^{\mathrm{ext}} + 2p \\
        \frac{f_1 p^2 v_I}{v_\mathrm{acc}^{2}} & \frac{f_1 p^2 v_Q}{v_\mathrm{acc}^{2}} \\
        \frac{f_2 p^2 v_Q}{v_\mathrm{acc}^{2}} & \frac{-f_2 p^2 v_I}{v_\mathrm{acc}^{2}}
    \end{bmatrix}
\end{equation}

\section{Discretization}\label{app:discretization}
For verification in simulation and experimental validation, the developed continuous model is discretized by zero-order hold with constant inputs between sampling instances.

By exploiting the Matrix exponential and by defining
\begin{align}
    \bar{A}_w \coloneq -\dfrac{1}{w}A &= \begin{bmatrix}
        \mathbf{I} & -\frac{1}{w}V\\ 0 & 0
    \end{bmatrix} \\
    \Rightarrow \quad \forall n\in\mathbb{N}\backslash\{0\}: \quad A^{n} &= (-w)^{n}\bar{A}_{w}
\end{align}
where $w \coloneqq \omega_{1/2}^\mathrm{ext}$ to improve readability, one obtains the discrete time system matrix
\begin{align}
        \Phi(T) = e^{AT}
        & =\sum_{k=0}^{\infty}\dfrac{T^{k}}{k!}A^{k}
        =\begin{bmatrix}
            \mathbf{I}&0\\0&\mathbf{I}
        \end{bmatrix}+\bar{A}_{w}\sum_{k=1}^{\infty}\dfrac{(-wT)^{k}}{k!}\\
        &=\begin{bmatrix}
            \mathbf{I}e^{-wT} & -\frac{1}{w}V(e^{-wT}-1)\\
            0 & \mathbf{I}
        \end{bmatrix}.
\end{align}
The discrete time input matrix results in 
\begin{align}
    \begin{split}
        \Gamma(T) &= \int_{0}^{T}e^{A\tau}Bd\tau
        =\begin{bmatrix}
            2\mathbf{I}(1-e^{-wT}) \\ 0
        \end{bmatrix}.
    \end{split}
\end{align}
By introducing $\alpha = 1 - e^{-wT}$,
\begin{align}
    \Phi(T) &= \begin{bmatrix}
        (1-\alpha)\mathbf{I} & \frac{\alpha}{w} V\\[0.2em]
        0&\mathbf{I}
    \end{bmatrix}\\
    \Gamma(T) &= 2\alpha \begin{bmatrix}
        \mathbf{I} \\ 0
    \end{bmatrix}.
\end{align}

\section{Discrete Time Pole Placement}\label{app:discretePoles}
Analogous to the continuous case, the discrete time observer gain
\begin{equation}
	\Lambda_k = \begin{bmatrix}
		(\alpha-1+\beta) \mathbf{I} \\
		W_k
	\end{bmatrix}
\end{equation}
is designed using $W$ as given in \eqref{eqn:W_mu}. The closed loop dynamic system matrix
\begin{equation}
	\Phi_{k,\mathrm{cl}} = \Phi_k + \Lambda_k C = \begin{bmatrix}
		\beta \mathbf{I} & \frac{\alpha}{w}V_k \\
		W_k & \mathbf{I}
	\end{bmatrix}
\end{equation}
and its characteristic polynomial
\begin{align}
	\begin{split}
		\det(\lambda\mathbf{I}_4 - \Phi_{\mathrm{cl}}) &=  \dfrac{\alpha^2}{w^{2}}\det\left((\lambda-\beta)(\lambda-1)\dfrac{w}{\alpha}\mathbf{I}-VW\right),
	\end{split}
\end{align}
where $\lambda \neq 1$, lead to closed loop eigenvalues
\begin{align}
	\lambda_j =& \dfrac{\beta+1}{2} \pm \sqrt{\dfrac{(\beta +1)^{2}}{4} - \beta + \dfrac{\alpha}{w}\mu_i}\\
    \text{where} \quad
    \mu_i =& \dfrac{w}{\alpha}(\lambda-\beta)(\lambda-1)
\end{align}
and again $i \in \{1, 2\}$ and $j \in \{1,2,3,4\}$.

Introducing the continuous to discrete time eigenvalue mapping $\rho = e^{pT}$, tuning factors $\varphi_i\in\mathbb{R}$, and setting
\begin{align}
	\begin{split}
		\beta &= 2 \rho - 1\\
		\mu_i &= \varphi_i\dfrac{w}{\alpha}\left(\beta-\rho^{2}\right)= -\varphi_i\dfrac{w}{\alpha}\left(1-\rho\right)^2
	\end{split}
\end{align}
leads to the discrete time eigenvalues
\begin{equation}
    \lambda_j=\rho \pm (1-\rho)\sqrt{1-\varphi_i}
\end{equation}
with the same tuning mechanism as the continuous eigenvalues in~\eqref{eqn:conttimeeigs}.

Again, with both $\varphi_i = 1$ all four $\lambda_j=\rho=e^{pT}$, i.e. the observers estimation error dynamics are critically damped if $p<0$. Also, the choice of $W$ allows association of $\varphi_1$ and $\varphi_2$ to the speed of estimation of $\Delta\omega_{1/2}$ and $\Delta\omega$, respectively. For stable observer dynamics, one has to ensure that $v_\mathrm{acc}^{2} > 0$ and $\lambda_j$ must remain strictly inside the unit circle. To fulfill the Nyquist criterion, one must select a $p$ according to
\begin{equation}\label{eqn:kappa}
    -\frac{\pi}{T} < p < -\kappa w <  0
\end{equation}
where $\kappa > 1$ is an arbitrary minimum continuous time pole relation factor.
For $\varphi_i > 1$, one obtains complex conjugated discrete time poles. For the eigenvalues to remain inside the unit circle, $\varphi_i$ must satisfy
\begin{align}
    \varphi_i & < \dfrac{2}{1-\rho}.
\end{align}
For $\varphi_i < 1$, one obtains two discrete time poles on the real axis. The pole farther left in the z-plane is not problematic due to positivity of the exponential function, and the upper bound on $p$ in \eqref{eqn:kappa} implies
\begin{equation}
    \varphi_i > 1-\left(\dfrac{e^{-\kappa wT} - \rho}{1 - \rho}\right)^{2}
\end{equation}
hence
\begin{equation}
    \max\left(1-\left(\dfrac{e^{-\kappa wT} - \rho}{1 - \rho}\right)^{2}, 0\right) < \varphi_i < \dfrac{2}{1-\rho}.
\end{equation}

Then, the parameter varying observer gain providing time invariant observer dynamics while neglecting parameter derivatives is
\begin{align}
	\begin{split}
		\Lambda(x) &=\begin{bmatrix}
			\alpha-2+2\rho & 0\\
			0 & \alpha-2+2\rho \\
			-\mu_1\frac{v_I}{v_\mathrm{acc}^{2}} &
			-\mu_1\frac{v_Q}{v_\mathrm{acc}^{2}} \\
			-\mu_2\frac{v_Q}{v_\mathrm{acc}^{2}} &
            \mu_2 \frac{v_I}{v_\mathrm{acc}^{2}}
		\end{bmatrix}
	\end{split}.
\end{align}

\section{Tuning factor comparison}
The tuning factors $f_i$ and $\varphi_i$ introduced in Appendix~\ref{app:LTV_observer} and \ref{app:discretePoles} can be used to move a pole pair away from $p$ or $\rho$ by the same euclidean distance in opposite directions, either vertically for values above one, or horizontally for values below. However, changes in $f_i$ and $\varphi_i$ are not equivalent, since the map $\rho = e^{pT}$ is nonlinear.

Equivalent expressions can be found iff $f\coloneq f_1=f_2$ and thus $\varphi \coloneq\varphi_1=\varphi_2$. For given $p$ , $f$, and $T$, the corresponding discrete time poles are obtained by setting
\begin{align}
    \rho &= e^{pT}\cos(pT\sqrt{f-1})
    \\
    \varphi &=1-\tan^{2}\left(pT\sqrt{f-1}\right).
\end{align}

\section{Beam estimation}\label{sec:appendix:beam}
When operating with an off-crest beam, the estimated detuning represents the frequency offset to the detuning that is minimizing the required forward power. Accordingly, the estimated bandwidth deviation reflects the amount of energy transferred to or extracted from the cavity by the beam. If the beam induced generator term is known, it can be subtracted from the measured forward signal before feeding it into the NLO. Thus, the beam becomes transparent to the NLO and estimations remain unpertubed. For quench detection purposes, the transparent approach is preferable, whereas beam perturbed estimation is advantageous to track an energy optimal cavity tune. In the following, the effects of the latter approach on the estimated quantities is analyzed.

A steady state complex RF drive $u_\mathrm{ss}$ leading to the steady state complex RF cavity voltage $v_\mathrm{ss}$ without beam is described by inserting \eqref{eqn:intro:omega12approx} and \eqref{eqn:derivation:omega12} into \eqref{eqn:statespace_general}, such that 
\begin{align}\label{eqn:app:steadystate}
    2u_\mathrm{ss} = (1+z_\mathrm{ss} +jy_\mathrm{ss})v_\mathrm{ss}
\end{align}
after converting to complex notation with the complex unit $j$. Here, \mbox{$z_\mathrm{ss}=\Delta\omega_{1/2}/\omega_{1/2}^\mathrm{ext}$} and $y_\mathrm{ss}=\Delta\omega/\omega_{1/2}^\mathrm{ext}$ are the apparent half bandwidth deviation and detuning, respectively. 

Assume an additional complex RF drive term $u_b$ compensates beam loading effects to maintain an accelerating voltage of $v_\mathrm{ss}$. Then, the apparent half bandwidth deviation and detuning must be adjusted to fulfill \eqref{eqn:app:steadystate}, leading to
\begin{equation}
    2(u_\mathrm{ss}+u_b) = \Big(1+(z_\mathrm{ss} +z_b)+j(y_\mathrm{ss}+y_b)\Big)v_\mathrm{ss}
\end{equation}
where $z_b$ and $z_b$ are beam induced changes in apparent parameters, and thus
\begin{equation}
    2u_b = (z_b+jy_b)v_\mathrm{ss}.
\end{equation}
The complex beam loading compensation voltage $u_b$ maintaining an accelerating gradient of $v_\mathrm{acc}$ therefore induces a change in apparent half bandwidth and detuning of
\begin{align}
\begin{split}
    \Delta\omega_{1/2,b} &= \omega_{1/2}^\mathrm{ext}z_b = 2\omega_{1/2}^\mathrm{ext}\operatorname{Re}(\dfrac{u_b}{v_\mathrm{ss}})\\
    \Delta\omega_{b} &= \omega_{1/2}^\mathrm{ext}y_b = 2\omega_{1/2}^\mathrm{ext}\operatorname{Im}(\dfrac{u_b}{v_\mathrm{ss}}).
\end{split}
\end{align}

\section{Python implementation}

\begin{lstlisting}[language=Python]
# Copyright 2024 A. Bellandi et al.
# Copyright 2025 B. Richter et al.
#
# Permission is hereby granted, free of charge, to any person
# obtaining a copy of this software and associated documentation
# files (the "Software"), to deal in the Software without
# restriction, including without limitation the rights to use,
# copy, modify, merge, publish, distribute, sublicense, and/or
# sell copies of the Software, and to permit persons to whom the
# Software is furnished to do so, subject to the following
# conditions:
#
# The above copyright notice and this permission notice shall be
# included in all copies or substantial portions of the Software.
#
# THE SOFTWARE IS PROVIDED "AS IS", WITHOUT WARRANTY OF ANY KIND,
# EXPRESS OR IMPLIED, INCLUDING BUT NOT LIMITED TO THE WARRANTIES
# OF MERCHANTABILITY, FITNESS FOR A PARTICULAR PURPOSE AND
# NONINFRINGEMENT. IN NO EVENT SHALL THE AUTHORS OR COPYRIGHT
# HOLDERS BE LIABLE FOR ANY CLAIM, DAMAGES OR OTHER LIABILITY,
# WHETHER IN AN ACTION OF CONTRACT, TORT OR OTHERWISE, ARISING
# FROM, OUT OF OR IN CONNECTION WITH THE SOFTWARE OR THE USE OR
# OTHER DEALINGS IN THE SOFTWARE.

"""
  Luenberger observer simulation routine.
  See: B. Richter et al. 'Estimation of superconducting cavity
  bandwidth and detuning using a Luenberger observer.'
"""

import numpy as np

def luenberger_observer(
    probe, forward, half_bandwidth_ext, detuning_init,
    sample_rate, observer_bandwidth, amplitude_threshold,
    bandwidth_gain_factor = 1, detuning_gain_factor = 1,
    I_init = 0.0, Q_init = 0.0):
  """Simulate the Luenberger observer.
  Internally, bandwidth and detuning estimation is done with
  normalization by the external half bandwidth.

  Args:
    probe (np.ndarray):
      2D array with two columns that represents the cavity probe
      signal trace in I/Q. The first column is the in-phase part
      and the second one is the quadrature.

    forward (np.ndarray):
      2D array with two columns that represents the cavity forward
      signal trace in I/Q. The first column is the in-phase part
      and the second one is the quadrature.

    half_bandwidth_ext (float):
      External half bandwidth of the cavity in Hz.

    detuning_init (float):
      Initial detuning in Hz.

    sample_rate (float):
      System sample rate in Hz.

    observer_bandwidth (float):
      Observer bandwidth in Hz.

    amplitude_threshold (float):
      Minimum absolute probe amplitude for which the adaptation of
      bandwidth and detuning is active.

    bandwidth_gain_factor (float, optional)
      Excess bandwidth estimation bandwidth adjustment.
      Defaults to 1.

    detuning_gain_factor (float, optional)
      Detuning estimation bandwidth adjustment.
      Defaults to 1.

    I_init (float, optional):
      Initial inphase RF observer state in MV.
      Defaults to 0.

    Q_init (float, optional):
      Initial quadrature RF observer state in MV.
      Defaults to 0.

  Returns:
    (np.ndarray):
      2D array with four columns of the estimated values.
      The first two columns represents the estimated I/Q signals
      for the probe.
      The third column represents the excess bandwidth. The total
      bandwidth results from adding the external half bandwidth
      with this column.
      The fourth column represents the detuning.
  """

  samples = forward.shape[0]
  T = 1. / sample_rate

  amplitude_square_minimum = amplitude_threshold**2

  ## Initialize discrete time LTV model. Eq. (12)
  sys_diagonal = np.exp(-2.0 * np.pi * half_bandwidth_ext * T)
  alpha = 1.0 - sys_diagonal

  Phi = np.zeros((4, 4))
  Phi[0, 0] = Phi[1, 1] = sys_diagonal
  Phi[2, 2] = Phi[3, 3] = 1.

  Gam = np.zeros((4, 2))
  Gam[0, 0] = Gam[1, 1] = 2. * alpha

  C = np.zeros((2, 4))
  C[0, 0] = C[1, 1] = 1.


  ## Initialize gain matrix. Eqs. (C6, C12)
  rho = np.exp(-2. * np.pi * observer_bandwidth * T)
  mu_0 = -(1-rho)**2 / alpha
  mu_1 = bandwidth_gain_factor * mu_0
  mu_2 = detuning_gain_factor * mu_0

  Lambda = np.zeros((4, 2))
  Lambda[0, 0] = Lambda[1, 1] = 2. * rho - 1. - sys_diagonal

  # Display tuning factor limits. Eq. (C11)
  kappa = 10
  print(1 - (
    (np.exp(-2*np.pi*kappa*half_bandwidth_ext*T)-rho)/
    (1 - rho)
    )**2, " < phi <", 2/(1-rho))

  ## Set initial estimated state vector.
  x_estimate = np.zeros((samples, 4))
  x_estimate[0, 0] = I_init
  x_estimate[0, 1] = Q_init
  x_estimate[0, 3] = detuning_init / half_bandwidth_ext


  ## Reserve memory for predicted state vector
  x_predict = np.zeros((4))

  for i in range(1, samples):

    # Prediction step

    ## Update the system matrix coefficients. Eq. (12)
    Phi[0, 2] = -alpha * x_estimate[i-1, 0]
    Phi[0, 3] = -alpha * x_estimate[i-1, 1]
    Phi[1, 2] = Phi[0, 3]
    Phi[1, 3] = -Phi[0, 2]

    ## State prediction
    x_predict = (np.dot(Phi, x_estimate[i-1, :]) +
                 np.dot(Gam, forward[i-1, :]))

    # Update step

    ## State estimation Eqs. (5, 6)
    prediction_error = probe[i, :] - np.dot(C, x_predict)
    x_estimate[i, :] = (x_predict -
      np.dot(Lambda, prediction_error))

    ## Update the observer gain coefficients. (C12)

    ampl_sq = x_estimate[i, 0]**2 + x_estimate[i, 1]**2

    ## Check square amplitude condition
    if ampl_sq > amplitude_square_minimum:
      bandwidth_coefficient = mu_1 / ampl_sq
      detuning_coefficient = mu_2 / ampl_sq
    else:
      # Disable detuning and bandwidth update
      bandwidth_coefficient = 0.0
      detuning_coefficient = 0.0

    ## Update the gain matrix. Eq. (12)
    Lambda[2, 0] = -bandwidth_coefficient * x_estimate[i, 0]
    Lambda[2, 1] = -bandwidth_coefficient * x_estimate[i, 1]
    Lambda[3, 0] = -detuning_coefficient * x_estimate[i, 1]
    Lambda[3, 1] = detuning_coefficient * x_estimate[i, 0]

  # Reverse the normalization.
  x_estimate[:, 2] *= half_bandwidth_ext
  x_estimate[:, 3] *= -half_bandwidth_ext
  return x_estimate
\end{lstlisting}

\bibliography{main.bib}

\end{document}